\documentclass[fleqn,usenatbib]{mnras}

\usepackage{newtxtext,newtxmath}

\usepackage[T1]{fontenc}
\usepackage{ae,aecompl}

\usepackage{soul}

\usepackage{graphicx}	
\usepackage{amsmath}	
\usepackage{amssymb}	
\usepackage{framed}

\usepackage{xcolor}

\title[HCO and CH$_3$ reactivity on interstellar grains]{Revisiting the reactivity between HCO and CH$_3$ on interstellar grain surfaces}

\author[J. Enrique-Romero et al.]{
J. Enrique-Romero,$^{1,2}$
S. {\'A}lvarez-Barcia,$^{3}$
F. J. Kolb,$^{3}$
A. Rimola,$^{2}$\thanks{E-mail: albert.rimola@uab.cat}
C. Ceccarelli,$^{1}$
\newauthor
N. Balucani,$^{1,4,5}$
J. Meisner,$^{3,6}$
P. Ugliengo,$^{7}$
T. Lamberts,$^{3,8}$
J. K{\"a}stner$^{3}$
\\
$^{1}$Univ. Grenoble Alpes, CNRS, Institut de
Plan\'etologie et d'Astrophysique de Grenoble (IPAG), 38000 Grenoble, France\\
$^{2}$Departament de Qu{\'i}mica, Universitat Aut{\`o}noma de Barcelona, 08193 Bellaterra, Catalonia, Spain\\
$^{3}$Institut f{\"u}r Theoretische Chemie, Universit{\"a}t Stuttgart,
Pfaffenwaldring 55, 70569 Stuttgart, Germany\\
$^{4}$Dipartamento di Chimica, Biologia e Biotecnologie, Universit{\`a} degli Studi di Perugia, Via Elce di Sotto 8, 06123 Perugia, Italy\\
$^{5}$Osservatorio Astrofisico di Arcetri, Largo E. Fermi 5, 50125 Firenze, Italy\\
$^{6}$Dep. of Chemistry and The PULSE Institute, Stanford University, Stanford, California 94305, USA and SLAC National Accelerator Laboratory, Menlo Park, California 94025, USA\\
$^{4}$Dipartamento di Chimica, Biologia e Biotecnologie, Universit{\`a} degli Studi di Perugia, Via Elce di Sotto 8, 06123 Perugia, Italy\\
$^{7}$Dipartamento di Chimica and Nanostructured Interfaces and Surfaces (NIS), Universit{\`a} degli Studi di Torino, Via P. Giuria 7, 10125 Torino, Italy\\
$^{8}$Leiden Institute of Chemistry, Gorleaus Laboratories, Leiden University, PO Box 9502, 2300 RA Leiden, The Netherlands\\
}

\date{Accepted XXX. Received YYY; in original form ZZZ}

\pubyear{2019}

\begin{document}
\label{firstpage}
\pagerange{\pageref{firstpage}--\pageref{lastpage}}
\maketitle

\begin{abstract}
Formation of interstellar complex organic molecules is currently thought to be dominated by the barrierless coupling between radicals on the interstellar icy grain surfaces. 
Previous standard DFT results on the reactivity between CH$_3$ and HCO on amorphous water surfaces, showed that formation of CH$_4$ + CO by H transfer from HCO to CH$_3$ assisted by water molecules of the ice was the dominant channel.
However, the adopted description of the electronic structure of the biradical (i.e., CH$_3$/HCO) system was inadequate (without the broken-symmetry (BS) approach). In this work, we revisit the original results by means of BS-DFT both in gas phase and with one water molecule simulating the role of the ice. Results indicate that adoption of BS-DFT is mandatory to describe properly biradical systems. In the presence of the single water molecule, the water-assisted H transfer exhibits a high energy barrier. In contrast, CH$_3$CHO formation is found to be barrierless. However, direct H transfer from HCO to CH$_3$ to give CO and CH$_4$ presents a very low energy barrier, hence being a potential competitive channel to the radical coupling and indicating, moreover, that the physical insights of the original work remain valid.
\end{abstract}




\section{Introduction}\label{sec:intro}

Interstellar complex organic molecules (iCOMs) are usually defined as compounds of 6--13 atoms in which at least one is C (\cite{herbst2009,ceccarelli2017,Herbst_2017}).  They are complex only from the astronomical point of view, while they are the simplest organic compounds according to terrestrial standards. Since terrestrial life is based on organic chemistry, the existence of iCOMs is of fundamental importance to ultimately understand the possible astrochemical origins of life. 

iCOMs are widespread in the Universe. They have been detected in a great variety of astrophysical objects like star forming regions (e.g. \cite{rubin1971};\cite{cazaux2003};\cite{Kahane_2013}; \cite{Mendoza2014}; \cite{LopezSepulcre2015ALMA}; \cite{belloche2017}; \cite{ligterink_alma-pils_2017}; \cite{mcguire_detection_2018}), in the circumstellar envelopes of AGB stars (\cite{cernicharo_vizier_2000}), shocked regions (\cite{arce_complex_2008}; \cite{Codella2017}; \cite{lefloch_l1157_2017}) and even in external galaxies (\cite{Muller2013}).
Despite their presence has been known for decades, how iCOMs are synthesized is still an open question and under debate (\cite{herbst2009}; \cite{caselli2012}; \cite{woods2013}; \cite{balucani2015}; \cite{butscher_radical_2017}; \cite{butscher2019}; \cite{fedoseev_experimental_2015}; \cite{enrique2016}; \cite{EnriqueRomero2019}; \cite{gal2017}; \cite{rivilla2017}; \cite{vasyunin2017}; \cite{rimola2018}; \cite{lamberts2019}).
Two different paradigms have been proposed: i) on the surfaces of grains (either during the cold prestellar or warmer collapse phase (e.g. \cite{garrod2006}; \cite{woods2013}; \cite{fedoseev_experimental_2015}; \cite{Oberg_2016}), and ii) through reactions in the gas phase (e.g. \cite{charnley1992}; \cite{balucani2015}; \cite{skouteris2018}). The first paradigm assumes that whenever two radicals (e.g., created by UV photon and/or cosmic rays incidences) are in close proximity (e.g., because of their diffusion) they can react to form iCOMs in a barrierless way. In the second one, iced simple hydrogenated molecules are released into the gas phase (e.g., due to thermal desorption), where they react to form iCOMs through gas-phase reactions. Interestingly, a review on the formation of iCOMs on interstellar grain surfaces investigated by means quantum chemical calculations has recently appeared (\cite{Zamirri_2019}).  

Currently, the ``on-surface'' paradigm is the scheme mostly adopted in astrochemical models.  However, a first theoretical study of the reactivity of HCO and CH$_3$ on an amorphous water surface (AWS), which is the bulk of the ices that envelope interstellar grains in cold objects, showed that the combination of these two radicals does not necessarily leads to the formation of the iCOM acetaldehyde (CH$_3$CHO) (\cite{enrique2016}). This unexpected result called for and was followed by other studies of different systems and with different computational methods.
First, \citet{rimola2018} and \citet{EnriqueRomero2019} studied the formation of formamide (NH$_2$CHO) and acetaldehyde by reactions between HCO and NH$_2$ and HCO and CH$_3$ on a AWS model by means of static quantum chemical calculations. Subsequently, \citet{lamberts2019} studied the formation of acetaldehyde by reaction between of HCO and CH$_3$ on a CO-pure ice model by means of \textit{ab initio} molecular dynamics (AIMD) simulations. The three works confirmed the main finding by \cite{enrique2016}, namely, that the reactivity between the radical pairs does not lead exclusively to the formation of the iCOMs, but formation of  CO + NH$_3$ and CO + CH$_4$ via direct H abstraction can also take place. In view of these results, formation of iCOMs via the barrierless radical-radical combination scheme needs still to be validated.

In this article, we aim to revise the first calculations carried out on the CH$_3$ + HCO system (\cite{enrique2016}) which were based on the standard DFT approach. Since then, it has become clear that an improved treatment of the radicals spins is necessary (\cite{rimola2018,EnriqueRomero2019}).
The article is organised as follows: in \S~\ref{sec:treatment} we review the treatment of the spins of a biradical system, in \S~\ref{sec:comput-details} we provide the details of the new computations carried out in this article, in \S~\ref{sec:results} we show the results and in \S~\ref{sec:conclusions} we discuss the conclusions.

\section{Why a better treatment of the biradical wave-function is needed}\label{sec:treatment}
In a previous work by some of us (\cite{enrique2016}), the reactivity between HCO and CH$_3$ in the gas phase and on AWS modelled by H$_2$O ice clusters was theoretically studied with standard DFT calculations. In the gas-phase model\footnote{We loosely use the term ``gas-phase'' to refer to systems where no water molecule are involved. The reader has to bear in mind that these reactions cannot take place in the ISM unless a third body (i.e. the grain) absorbs the released nascent energy.}, different synthetic channels were identified, namely, the formation of acetaldehyde (CH$_3$CHO), CO + CH$_4$ and CH$_3$OCH, the occurrence of which being determined by the relative orientation of the radicals. In contrast, on the AWS models, a hydrogen-atom relay mechanism assisted by water molecules of the ice led to the exclusive formation of CO + CH$_4$.

The electronic ground state for the CH$_3$CHO, CO + CH$_4$ and CH$_3$OCH products is a singlet wave function as they are closed-shell systems. Conversely, the HCO and CH$_3$ radicals are open-shell doublet systems due to their unpaired electron, while a system consisting of the two radicals (i.e., HCO and CH$_3$ together) can be either in triplet or singlet electronic states (the spins of the unpaired electrons can be of the same sign or of opposite signs, respectively). The triplet state is electronically non-reactive due to the Pauli repulsion. In contrast, the singlet state (usually referred to as a biradical system) is reactive because of the opposite spin signs. The description of the electronic structure of biradical states requires a wave function composed of more than one Slater determinant to recover static correlation. In the wave function-based post-Hartree-Fock (post-HF) realm, this can be described by multi-configurational self consistent field (MCSCF) methods, such as the complete active space self-consistent field (CASSCF), or the so-called multi-reference methods like the complete active space perturbation theory (CASPTn) ones. In CASSCF, a particular number of electrons ($N$) are distributed between all possible (namely, ground and excited) configurations that can be constructed from $M$ molecular orbitals, i.e., an ($N,M$) active space. It is worth mentioning that one has to pay special care when deciding the orbitals to include in the active space, since the resulting wave-function could erroneously describe the system under study. CASPTn is an improvement over CASSCF($N,M$) where a perturbative expansion is further performed in order to retrieve more dynamic electron correlation. On the other hand, such multi reference character cannot be obtained from normal Kohn-Sham density functional theory (DFT). Instead, the electronic structure of biradicals can be approximated by an unrestricted open-shell wave function with the broken-(spin)-symmetry \textit{ansatz}, where a triplet state is mixed with a combination of ground and excited singlet states in order to obtain an electron-correlated wave-function (\cite{noodleman_valence_1981}, \cite{noodleman_electronic_1984}, \cite{neese_definition_2004}). 
%

Calculations by \cite{enrique2016} were performed in an open-shell formalism, but after publication we realised that the initial guess wave functions remained in a metastable, symmetric state with spin-up and spin-down orbitals being equally mixed (i.e., spin analysis indicated 50\% of spin up and 50\% of spin down in both radicals and the total spin density being  zero), thus resembling  a closed-shell solution. Compared to that, the actual broken-symmetry wave function leads to a significant stabilization of the reactants, which changes the results qualitatively. Thus, the present work aims to  revise some of the original results using the DFT broken-symmetry solution, showing moreover that it agrees reasonably well with those at the CASPT2 level.

\section{Computational details}\label{sec:comput-details}
All DFT calculations were performed using the GAUSSIAN09 package (\cite{gaussian09}), while post-HF multi-configurational and multi-reference calculations were carried out with the OpenMolcas 18.09 software (\cite{Molcas}, \cite{Molcas7}, \cite{Molcas8}, \cite{OpenMolcas}).

DFT geometry optimisations and transition state searches  were carried out with i) the M06-2X (\cite{M06ZahoTruhlar}) and ii) BHLYP-D3 (i.e. BHLYP, \cite{bhandhlyp-becke1993}, \cite{lee1988}) including the Grimme's D3 dispersion correction (\cite{D2-grimme2006}, \cite{D3-grimme2010}) functionals, in combination with a def2-TZVPD basis set. Structures with triplet electronic states were simulated with open-shell calculations based on an unrestricted  formalism. Singlet biradical systems were calculated adopting an unrestricted broken-symmetry (BS) approach. For the sake of comparison, for some cases, single point energy calculations adopting standard (i.e., non-BS) unrestricted (U) formalisms  have also been carried out.

CASSCF geometry optimisations and transition state searches were performed using a (2,2) active space, corresponding to the radical unpaired electrons in their respective orbitals. Reaction energetics were refined by performing CASPT2 single point energy calculations on the CASSCF(2,2) optimised geometries. In both cases the cc-pVDZ basis set was employed. For the sake of clarity, here we only show the CASPT2 results, the CASSCF(2,2) ones being available as on-line supporting information (SI). 

Since the scope of this work is to revise the electronic structure of the biradical systems, only electronic energy values are reported and accordingly zero-point energy corrections were not accounted for here.

Input file examples for this kind of calculations are provided in SI.\\

\section{Results}\label{sec:results}


\subsection{Reactions in the gas phase}
\begin{figure}
    \centering
    \includegraphics[width=0.47\textwidth]{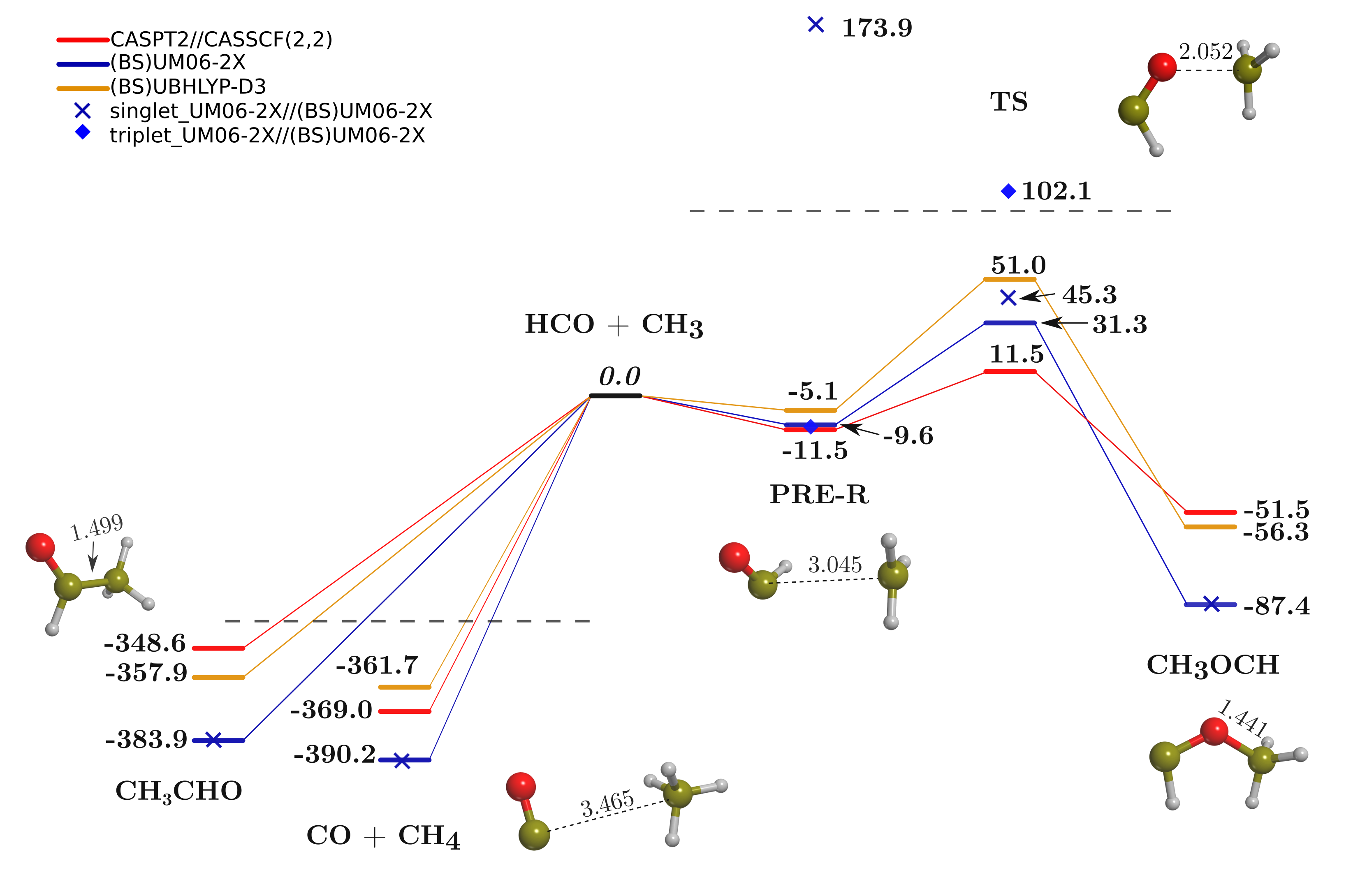}
    \caption{PESs at different DFT levels and at CASPT2 for the reactivity between CH$_3$ and HCO to form CH$_3$OCH (right side panel) or CO + CH$_4$ and CH$_3$CHO (left side panel). The energy reference 0.0 is the HCO + CH$_3$ asymptote. Dashed horizontal lines indicate broken vertical axis. PRE-R refers to the prereactant complexes and TS to the transition states. Single point energies at singlet and triplet UM06-2X levels on the (BS)UM06-2X optimised geometries are also shown. The presented structures correspond to the (BS)UM06-2X optimized geometries except for PRE-R, wich is the triplet UM06-2X optimized geometry. Energy units are in kJ mol$^{-1}$ and distances in \AA. We have also checked the triplet state of the CH$_3$OCH product resulting 80 kJ mol$^{-1}$ higher in energy than the singlet state and an energy barrier for its formation about 55 kJ mol$^{-1}$ higher than the singlet case, (UM06-2X theory level).}
    \label{fig:CO_gas}
\end{figure}

In the gas phase, in agreement to the previous work \cite{enrique2016}, the nature of the final product depends on the relative initial orientations of the reactants. When the C atoms of the two radicals are pointing one to each other (i.e. H$_3$C$\cdots$CHO), they couple to form CH$_3$CHO, similarly, when the H atom of the HCO is pointing to the C atom of CH$_3$ (i.e. OCH$\cdots$CH$_3$), H is transferred to form CO + CH$_4$. Both processes have been found to be barrier-less, irrespective of the method (i.e., BS-DFT and MCSCF, see the left hand side of Figure \ref{fig:CO_gas}).

On the other hand, when the O atom of HCO points towards the C atom of CH$_3$ (i.e. HCO$\cdots$CH$_3$), the carbene CH$_3$OCH species can form. For this case, however, BS-DFT calculations indicate that the biradical system is metastable. Consequently, formation of CH$_3$OCH is not spontaneous but requires overcoming an energy barrier of 31.3 and 51.0 kJ mol$^{-1}$ at the (BS)UM06-2X and (BS)UBHLYP-D3 levels, respectively. The same trend is found for CASPT2 calculations with an energy barrier of 11.5 kJ mol$^{-1}$ (see right hand side of Figure \ref{fig:CO_gas}).
It is worth mentioning that, for the formation of CH$_3$OCH, U single point calculations on the (BS)UM06-2X optimized geometries (without considering the BS approach) result in spontaneous formation of CH$_3$OCH, leading to the same result as for the restricted situation (see singlet\_UM06-2X energies in Figure \ref{fig:CO_gas} represented by blue crosses). This is because the singlet UM06-2X initial guess wave function does not consider the reactant as an actual biradical system but the unpaired electrons are localized 50\% spin-up and 50\% spin-down in one radical and the same for the other radical, resembling an electronic closed-shell situation. This excited initial guess wave function is about 173.9 kJ mol$^{-1}$ less stable than the asymptote (0.0 kJ mol$^{-1}$, corresponding to the situation where the radicals are infinitely separated) and hence the system rolls down to the most stable closed-shell situation. 
Similarly, single points at the triplet UM06-2X level on the (BS)UM06-2X optimized geometries are also shown in Figure \ref{fig:CO_gas} (represented by blue diamonds). We want to stress out that triplet-state wave-functions do not require the use of the broken-symmetry \textit{ansatz} as single-reference methods like UDFT already provide good descriptions of such open shell systems thanks to Pauli's exclusion principle.

\subsection{Reactions in the presence of one water molecule}

For the reactivity between CH$_3$ and HCO in the presence of one water molecule, we have studied the reactions of CH$_3$CHO formation through a radical-radical coupling (Rc) and the formation of CO + CH$_4$ through both a direct hydrogen abstraction (dHa), i.e., the H transfer is direct from HCO to CH$_3$, and a water-assisted hydrogen transfer (wHt), i.e., the H transfer is assisted by the water molecule which allows a successive H-transfer mechanism OC$\cdot\cdot$H$\cdots$HO$\cdot\cdot$H$\cdots$CH$_3$.

In \cite{enrique2016} it was shown that, in the presence of (H$_2$O)$_{18}$ and (H$_2$O)$_{33}$ water cluster models, the wHt was found to be barrierless, i.e., the {\bf assisted} H transfer occurred spontaneously during the optimization process, a finding that led the authors to conclude that this channel was the dominant one over the others. However, we identified that such a spontaneous process is a consequence of the limitation of standard DFT to describe the electronic structure of biradical systems if the BS approach is not adopted.

\begin{figure}
    \centering
    \includegraphics[width=0.45\textwidth]{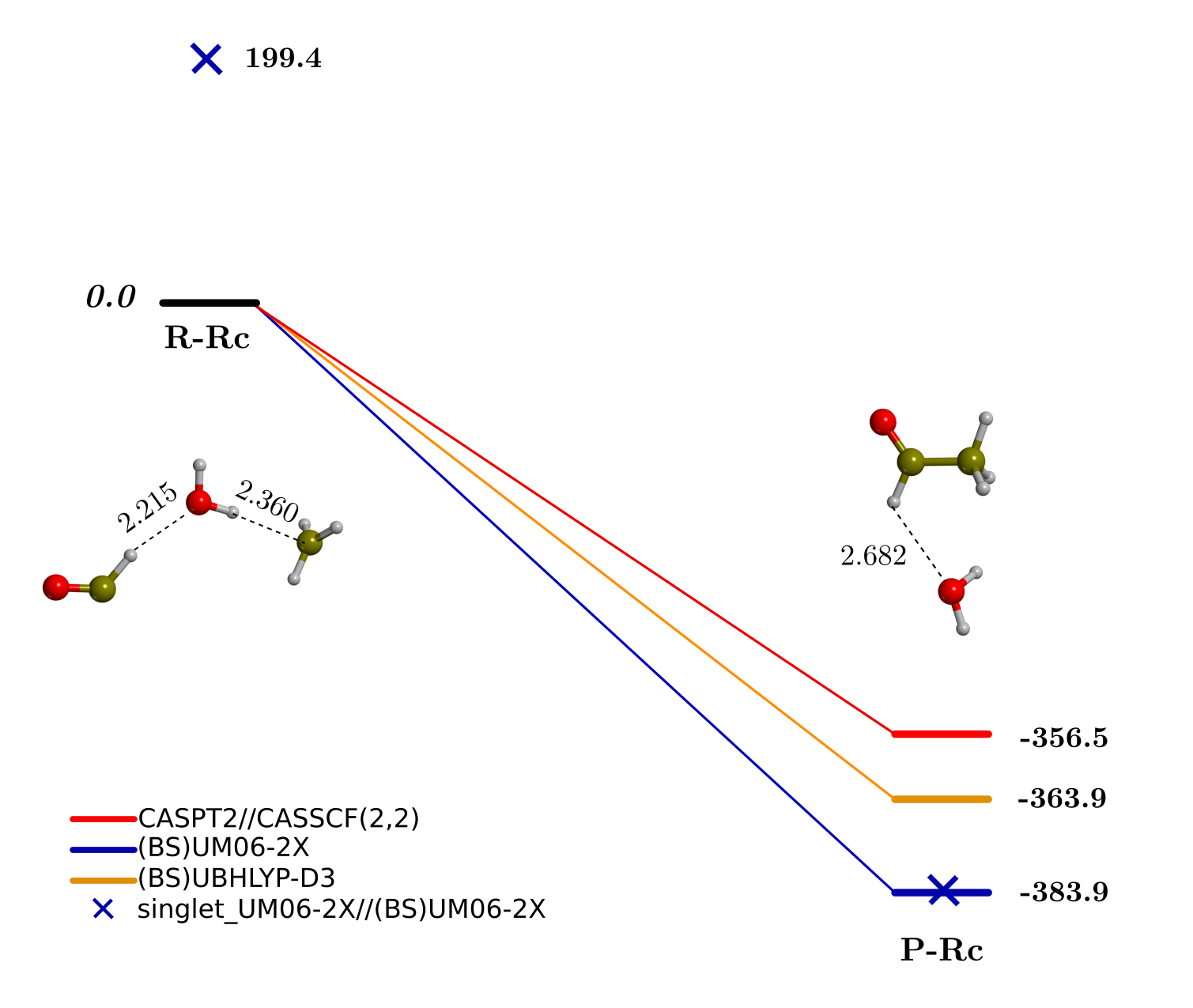}
    \caption{PESs at different DFT levels and at CASPT2 for the Rc reaction between CH$_3$ and HCO in the presence of one water molecule. The 0th energy reference correspond to the prereactant complex (R-Rc). Single point energies at the singlet UM06-2X UM06-2X level on the (BS)UM06-2X optimised geometries are also shown. The structures presented in this figure are those corresponding to the (BS)UM06-2X optimised geometries. Energy units are in kJ mol$^{-1}$ and distances in \AA.}
    \label{fig:Rc}
\end{figure}

By adopting BS-DFT, we have found here that for the both (BS)UM06-2X and (BS)UBHLYP-D3 methods, Rc is a barrierless process (see Figure \ref{fig:Rc}).

\begin{figure}
    \centering
    \includegraphics[width=0.45\textwidth]{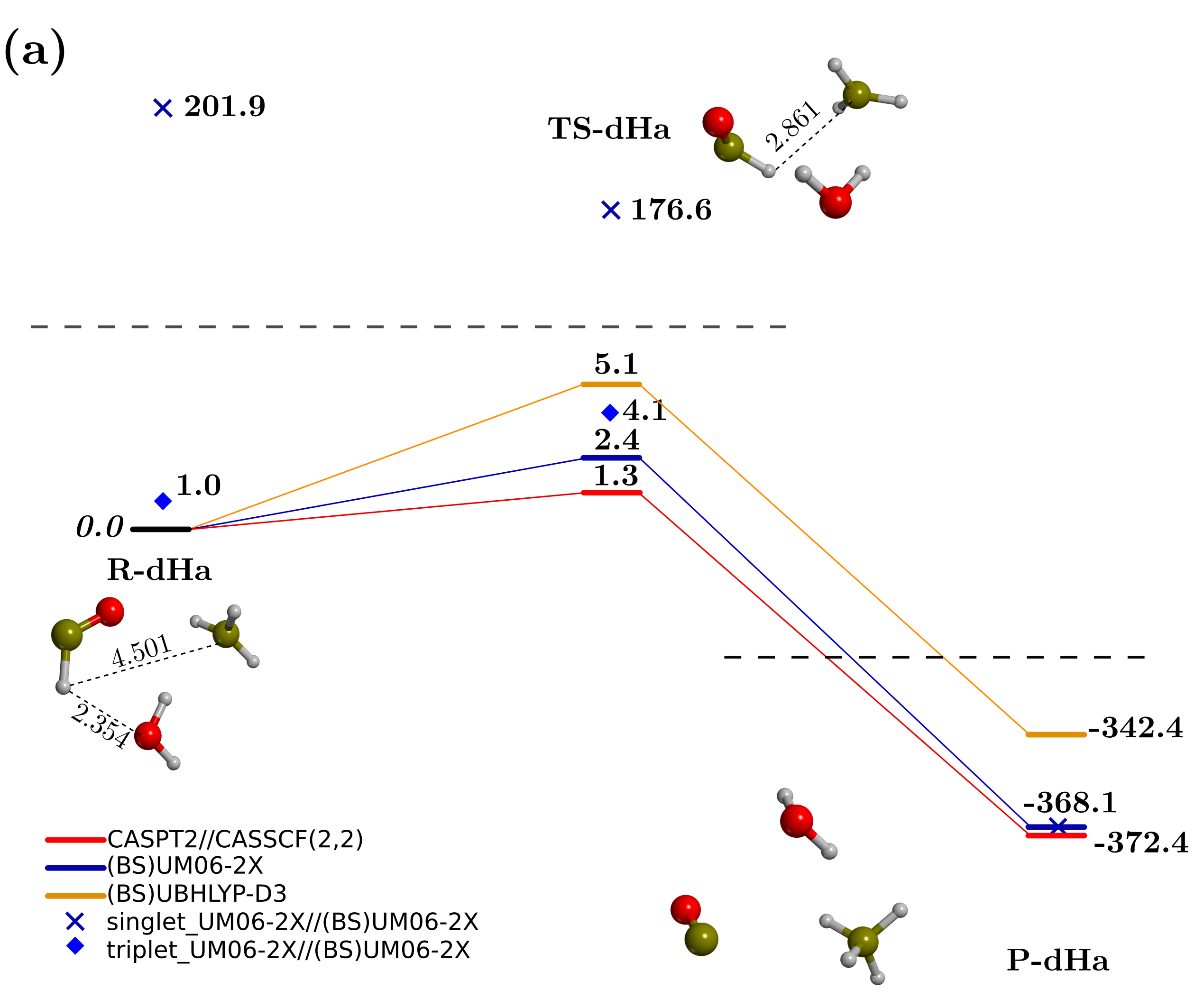}
    \includegraphics[width=0.45\textwidth]{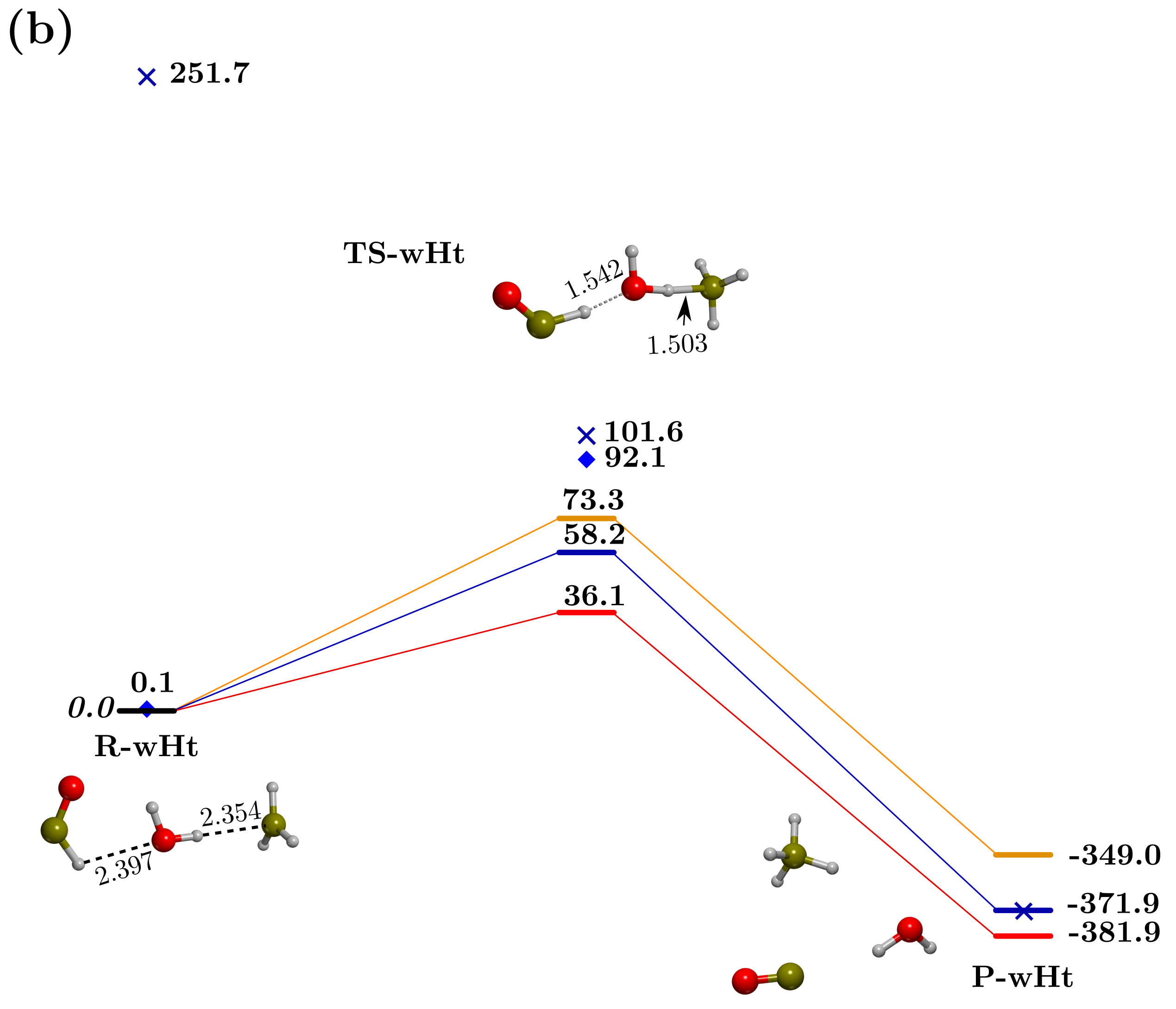}  
    \caption{PESs at different DFT levels and at CASPT2 for the reactions of dHa (a) and wHt (b) between CH$_3$ and HCO in the presence of one water molecule. The 0th energy references are the prereactant complexes: R-dHA (a) and R-wHt (b). Single point energies at the singlet UM06-2X and triplet UM06-2X levels on the (BS)UM06-2X optimised geometries are also shown. Dashed horizontal lines indicate broken vertical axis. The presented structures correspond to the (BS)UM06-2X optimised geometries. Energy units are in kJ mol$^{-1}$ and distances in \AA.}
    \label{fig:dHa-wHt}
\end{figure}

The PESs for the dHa and wHt processes at the different theory levels are shown in Figure \ref{fig:dHa-wHt}. At (BS)UM06-2X and (BS)UBHLYP-D3 levels, dHa presents a small energy barrier (2.4 and 5.1 kJ mol$^{-1}$, respectively). In contrast, wHt presents a high energy barrier (58.2 and 73.3 kJ mol$^{-1}$, respectively), indicating that it is not spontaneous. Similar findings are provided by CASPT2, which predicts energy barriers of 1.3 and 36.1 kJ mol$^{-1}$ for dHa and wHt, respectively. In contrast, U single point energy calculations on the (BS)UM06-2X optimized geometries without considering the BS approach describe both dHa and wHt as spontaneous processes (see singlet\_UM06-2X energies in Figure \ref{fig:dHa-wHt} represented by blue crosses), in which the reactant structures lay above the actual reactants by more than 200 and 250 kJ mol$^{-1}$, respectively. This is because the singlet UM06-2X calculation starts from a non-symmetry broken initial guess wave function, hence yielding the same wave function as a restricted (i.e., closed-shell) M06-2X calculation. This calculated closed-shell wave function can be understood as an electronically excited state, in which the electronic structure has a signifcant contribution of an ionic state: a protonated CO molecule (HCO$^+$) and a negatively charged CH$_3$ species (CH$_3^-$). This ionic state is an ideal situation to trigger a Grotthus-like mechanism, in which the ``extra'' proton of HCO$^+$ is transferred through the assisting water molecule to the ``proton-defective''  CH$_3^-$.
These results  confirm again the need to use the BS-DFT approach to properly describe biradical systems.

\section{Conclusions}\label{sec:conclusions}

This work is a revision note of a previous work by some of us \cite{enrique2016}, in which the reactivity of the same system, i.e., CH$_3$ + HCO, has been studied using DFT methods adopting an unrestricted broken-symmetry approach (i.e., (BS)UM06-2X and (BS)UBHLYP-D3) as well as post-HF multi-configurational and multi-reference methods (i.e., CASSCF(2,2) and CASPT2). In the original work, the DFT broken-symmetry formalism was not adopted, hence seriously affecting the description of the electronic structure of the CH$_3$/HCO biradical system. The main conclusions of the present work are summarized as follows:
\begin{itemize}
    \item When the unrestricted DFT formalism is used without adopting the broken-symmetry approach to describe the electronic structure of biradical systems, the initial guess wave function may collapse into a restricted closed-shell solution. If this occurs, the reactivity between two radicals is likely to be wrongly described. In the particular case of the CH$_3$ + HCO reactivity on water ice,  calculations indicate that the water assisted H transfer process is spontaneous.
    \item Unrestricted broken-symmetry DFT calculations for biradical systems show qualitatively similar results as those obtained at post-HF multi-configurational and multi-reference levels, indicating the suitability of this DFT approach to describe the reactivity of biradical systems. 
    \item In the gas phase, both CH$_3$CHO and  CO + CH$_4$ formations are found to be barrierless. In contrast, formation of the carbene CH$_3$OCH species has a noticeable barrier.
    \item In the presence of one water molecule, the water assisted hydrogen transfer reaction is not spontaneous but, in contrast, it has a high energy (58 and 73 kJ mol$^{-1}$ at the (BS)M06-2X and (BS)BHLYP-D3 levels). Accordingly, its occurrence is unlikely under the interstellar conditions. In contrast, the radical-radical coupling is barrierless and the direct hydrogen abstraction presents a very small energy barrier (5 kJ mol$^{-1}$ at the most). Similar results have been obtained using larger cluster models mimicking the surface of interstellar water ice \cite{EnriqueRomero2019}. 
 \end{itemize}

Finally, it is worth mentioning that, despite the limited description of the biradical system in \cite{enrique2016}, the physical insights provided by that work remain still valid, since it is shown that the biradical reactivity does not necessarily result in the radical-radical coupling product (i.e., the iCOM). Indeed, it is found here that the direct hydrogen abstraction can actually be a competitive channel, giving the same product as that for the water assisted hydrogen transfer. This finding is in agreement with recent theoretical works dealing with the reactivity of biradical systems on interstellar ice surfaces (\cite{rimola2018, EnriqueRomero2019, lamberts2019}).

\section*{Acknowledgements}

JER and CC acknowledges funding from the European Research Council (ERC) under the European Union's Horizon 2020 research and innovation program, for the Project ``the Dawn of Organic Chemistry" (DOC), grant agreement No 741002. AR is indebted to the ``Ram{\'o}n y Cajal" program. MINECO (project CTQ2017-89132-P) and DIUE (project 2017SGR1323) are acknowledged. TL and JK acknowledge financial supported by the European Union's Horizon 2020 research  and  innovation  programme  (grant  agreement  No. 646717,  TUNNELCHEM),  the  Alexander  von  Humboldt Foundation, and the Netherlands Organisation for Scientific Research (NWO) via a VENI fellowship (722.017.008). We additionally ackowledge the GRICAD infrastructure (https://gricad.univ-grenoble-alpes.fr), which is partly supported by the Equip@Meso project (reference ANR-10-EQPX-29-01) of the programme Investissements d'Avenir supervised by the Agence Nationale pour la Recherche and the HPC resources of IDRIS under the allocation 2019-A0060810797 attributed by GENCI (Grand Equipement National de Calcul Intensif).
PU and NB acknowledge MIUR (Ministero dellIstruzione, dellUniversità e della Ricerca) and Scuola Normale Superiore (project PRIN 2015, STARS in the CAOS - Simulation Tools for Astrochemical Reactivity and Spectroscopy in the Cyberinfrastructure for Astrochemical Organic Species, cod. 2015F59J3R). This project has received funding from the European Unions Horizon 2020 research and innovation programme under the Marie Skodowska-Curie grant agreement No 811312




\bibliographystyle{mnras}
\bibliography{mybib} 




\appendix

\section{On-line material}

In the on-line material file we provide: i) The absolute energies (in Hartrees) of the reactions presented in this work, ii) some input examples for Gaussian and OpenMolcas calculations, and iii) the XYZ Cartesian coordinates of the structures presented in the work.


\bsp	
\label{lastpage}
\end{document}